\documentclass[a4paper]{article}
\pdfoutput=1
\usepackage{listings}
\usepackage{xcolor}
\usepackage{multirow}
\usepackage{jheppub}  
\usepackage{dcolumn}
\usepackage[english]{babel}
\usepackage[utf8x]{inputenc}
\usepackage[T1]{fontenc}
\usepackage{palatino}
\usepackage{amsmath}
\usepackage{afterpage}
\usepackage{graphicx, subcaption}
\usepackage[colorinlistoftodos]{todonotes}
\usepackage[colorlinks=true]{hyperref}
\usepackage{makeidx}
\newcommand{\be}{\begin{equation}}  
\newcommand{\ee}{\end{equation}}  
\newcommand{\bea}{\begin{eqnarray}}  
\newcommand{\eea}{\end{eqnarray}}  
\makeindex
\begin{document}

\vspace*{1.2cm}

\thispagestyle{empty}
\begin{center}
{\LARGE \bf Two-particle Correlations in multi-Regge Kinematics }

\par\vspace*{7mm}\par

{

\bigskip

\large \bf N. Bethencourt de Le\'on$^1$, G. Chachamis$^2$, A. Sabio Vera$^{1,3}$}

\bigskip

{\large \bf  E-Mail: chachamis@gmail.com}

\bigskip

{$^1$ Instituto de F{\'\i}sica Te{\'o}rica UAM/CSIC, Nicol{\'a}s Cabrera 15, E-28049 Madrid, Spain.\\
$^2$ Laborat{\' o}rio de Instrumenta\c{c}{\~ a}o e F{\' \i}sica Experimental de Part{\' \i}culas (LIP),\\
Av. Prof. Gama Pinto, 2, P-1649-003 Lisboa, Portugal.\\
$^3$ Theoretical Physics Department, Universidad Aut{\' o}noma de Madrid, E-28049 Madrid, Spain.\\ }

\bigskip

{\it Presented at the Low-$x$ Workshop, Elba Island, Italy, September 27--October 1 2021}

\vspace*{15mm}

\end{center}
\vspace*{-5mm}

\begin{center}
{\bf Abstract}
\end{center}
\begin{abstract}.
Multi-jet production at the LHC is an important  process to study. In particular, events with final state kinematic configurations where we have two jets widely separated in rapidity with similar $p_T$ and lots of mini-jets or jets populating the space in between are relevant for the high energy limit of QCD.
Keeping the jet multiplicity fixed, the study of these events is a good ground to test different models of multi-particle production in hadron-hadron collisions. We report on a comparison between the predictions of the old multiperipheral Chew-Pignotti model and those of BFKL for the single jet rapidity distributions and for jet-jet rapidity correlations.
\end{abstract}
 
 \section{Introduction}

An important question in Quantum Chromodynamics (QCD) is the structure of the high energy limit dynamics of multi-particle production at hadron and hadron-lepton colliders. Multi-particle events, and in particular, multi-jet
events, are difficult to describe when one wants to go beyond the leading order (LO) approximation in fixed order perturbation theory. Even in the early days of hadron colliders (Intersecting Storage Rings, ISR at CERN) when 
the colliding energy was of the order of few tens of GeV and QCD was not yet the established theory of the strong interaction, multi-particle production was one of the key problems to tackle in order to probe the underlying dynamics
even on heuristic terms.  The experimental data amassed in the last 50-60 years, starting from those early attempts,  
have generally reinforced the view that the final state particles (pions, kaons, protons, etc) that end up on the 
detector seem to belong in correlated ``clusters''~\cite{Dremin:1977wc} that are emitted in the hard scattering part
of the process or on the partonic level. 

Our modern day picture of a generic partonic cross section
of two incoming partons that interact and produce two or more outgoing partons which in turn,
through a parton shower, radiate new quarks and gluons and form jets does not contradict 
the heuristic notion of  clusters.
The study of differential distributions and particle-particle correlations between final states had given
important insights about the strong interaction no matter whether
 the  final states were considered before hadronizations effects (final state partons/jets) or after (hadrons in the detector calorimeters).
Especially for jets, correlations summarize important properties without being too sensitive
to unresolved soft particles in the jet~\cite{Tannenbaum:2005by} (see also the introduction of Ref.~\cite{SanchisLozano:2008te}). It is thus very interesting to see whether the rapidity distributions of final state hadrons are any similar to the rapidity distributions of jets.

Quite a few different approaches exist that are useful to probe multi-particle hadroproduction beyond fixed order. These are  resummation frameworks that resum different leading contributions to amplitudes (e.g. DGLAP~\cite{Gribov:1972ri,Gribov:1972rt,Lipatov:1974qm,Altarelli:1977zs,Dokshitzer:1977sg}, BFKL~\cite{Kuraev:1977fs,Kuraev:1976ge,Fadin:1975cb,Lipatov:1976zz,Balitsky:1978ic,Lipatov:1985uk}, CCFM~\cite{Ciafaloni:1987ur,Catani:1989yc}, Linked Dipole Chain~\cite{Gustafson:1986db,Gustafson:1987rq,Andersson:1995ju}, Lund Model~\cite{Andersson:1998tv}, see also~\cite{LHCForwardPhysicsWorkingGroup:2016ote}). 

The run 2 of the LHC at 13 TeV has provided lots of data which should be analyzed in detail. Here, we want to focus on a final state configurations which could be quite important to answer the question of what
is the range of applicability of different multi-particle production models. These configurations correspond to events with several final state jets where the two outermost in rapidity ones are also well separated in rapidity and can be  tagged requiring that their transverse momenta are similar and genaraly large. These corresponds to a subset of the so-called Mueller-Navelet jets~\cite{Mueller:1986ey} and if additionally we require a very similar $p_T$ for the outermost jets (however, not within overlapping ranges) we can enforce that their influence on 
the rapidity distributions of the jets in between will be symmetric.

Our aim here
is to report on our recent work~\cite{deLeon:2020myv} where we studied some of the characteristics that can be attributed to these processes. As the most striking features in multiperipheral models can be found in the rapidity space --the origin of these features can be traced at the decoupling of the transverse degrees of freedom from the longitudinal ones--, we will present single rapidity differential distributions and two-jet rapidity-rapidity correlations. 

Assuming configurations in a hadron-hadron collider with fixed final state jet multiplicity $N$,
the single and double rapidity distributions for a given jet and a pair of jets are given respectively by
$\rho_{1}(y_i)$ and $\rho_{2}\left(y_{i}, y_{j}\right)$, where $i$ and $j$ are the positions of the jets
once they are ordered in rapidity with $1 \le i,j \le N$. We can formally define for the 
single and double differential normalized distributions
\begin{equation}
\rho_{1}(y_i)=\frac{1}{\sigma} \int d^{2} p_{\perp} \frac{d^{3} \sigma}{d y_i d^{2} p_{\perp i}}
\end{equation}
 and
\begin{equation}
\rho_{2}\left(y_{i}, y_{j}\right)=\frac{1}{\sigma} \int d^{2} p_{\perp i} d^{2} p_{\perp j} \frac{d^{6} \sigma}{d y_{i} d^{2} p_{\perp i} d y_{j} d^{2} p_{\perp j}}\,.
\end{equation}
After integrating over their transverse components we will have
\begin{equation}
\rho_{1}(y_i)=\frac{1}{\sigma} \frac{d \sigma}{d y_i}
\end{equation}
 and
\begin{equation}
\rho_{2}\left(y_{i}, y_{j}\right)=\frac{1}{\sigma}  \frac{d^{2} \sigma}{d y_{i}  d y_{j} }\,.
\end{equation}
The two-particle rapidity-rapidity correlation function is then given by
\begin{eqnarray}
C_{2}\left(y_{i},  y_{j}\right)&=&\frac{1}{\sigma} \frac{d^{2} \sigma}{d y_{i} d y_{j} }-\frac{1}{\sigma^{2}} \frac{d \sigma}{d y_{i} } \frac{d \sigma}{d y_{j}}\\
&=& \rho_{2}\left(y_{i}, y_{j}\right) - \rho_{1}(y_i)  \rho_{1}(y_j)
\end{eqnarray}
In practice however, the correlation function is usually computed with the following expression which is less sensitive
to experimental errors:
\begin{equation}
R_{2}\left(y_{1}, y_{2}\right)=\frac{C_{2}\left(y_{1}, y_{2}\right)}{\rho_{1}\left(y_{1}\right) \rho_{1}\left(y_{2}\right)}=\frac{\rho_{2}\left(y_{1}, y_{2}\right)}{\rho_{1}\left(y_{1}\right) \rho_{1}\left(y_{2}\right)}-1\,.
\label{correlation_function}
\end{equation}

In Section 2, we calculate the analytic expressions for the double and single rapidity distributions within an
old multiperipheral model, namely the Chew-Pignotti model~\cite{Chew:1968fe}, wehreas, In Section 3,
we explain how we compute the same quantities for the gluon Green's function of a collinear BFKL model by
using Monte Carlo techniques. We conclude in Section 4. 
\section{The Chew-Pignotti Model}

As we mentioned in the introduction, we will work with a Chew-Pignotti type of multiperipheral model~\cite{Chew:1968fe} following the analysis by DeTar in Ref.~\cite{Detar:1971qw}. The simplified features of the model
 will give us the opportunity to produce analytic results that could in principle
 be compared to the experimental data. For a more detailed review on multiperipheral models and the cluster concept in multiparticle production at hadron colliders, see Ref.~\cite{Dremin:1977wc} and references therein. 
The key point is that in these types of models, the transverse coordinates decouple from the longitudinal degrees
of freedom (rapidity) and that is what allows to obtain analytic expressions for the rapidity distributions.
 
 While the multiperipheral models were devised to describe multiple particle production, we will use
 the Chew-Pignotti here for multiple jet production. We will assume that the rapidity separation of
 the the bounding jets is Y and also that the jets in between have a fixed multiplicity $N$ such that in total
 we will have $N+2$ final state jets in each event. The outermost in rapidity jets (the most forward/backward jets)  are having rapidities $\pm \frac{Y}{2}$. It should be clear that we choose to work with limits $y_0=-\frac{Y}{2}$ and $y_{N+1}=\frac{Y}{2}$ because we want to cast our analytic expressions for the distributions in a symmetric
 way with respect to the forward and backward rapidity direction. Naturally, one could also work with limits $y_0=0$ and $y_{N+1}=Y$. Actually for the results from the BFKL approach in section 3,  we choose to present our plots in
the range $\left[0, Y \right]$ as we want to be closer to an experimental analysis setup.

The cross section for the production of  $N+2$ final state jets is given by
\begin{eqnarray}
\sigma_{N+2} &=& \alpha^{N+2} \int_{0}^{Y} \prod_{i=1}^{N+1} dz_i \delta \left(Y-
\sum_{s=1}^{N+1} z_s \right) \nonumber\\
&=&  \alpha^{N+2}
\int_{-\frac{Y}{2}}^{\frac{Y}{2}} dy_N \int_{-\frac{Y}{2}}^{y_N} dy_{N-1} \cdots \int_{-\frac{Y}{2}}^{y_3} dy_2 
\int_{-\frac{Y}{2}}^{y_2} dy_1\, = \,  \alpha^2 
\frac{\left(\alpha Y\right)^N}{N!} \, ,
\label{sigma_N2}
\end{eqnarray}
which leads to a total cross section $\sigma_{\rm total} = 
\sum_{N=0}^\infty \sigma_{N+2} = \alpha^2 e^{\alpha Y}$ and the rise with $Y$ would need to be tamed
by introducing unitarity corrections in transverse space. A rapidity 
$y_l$, with $l=0, \dots, N+1$ is assigned to each of the final-state jets. At $y_0=-\frac{Y}{2}$ and $y_{N+1}=\frac{Y}{2}$ we have the positions of the outermost jets with the jet vertex reduced to a simple factor equal to $\alpha$, the strong coupling constant. The in-between jets will have rapidities $y_l = -\frac{Y}{2}+\sum_{j=1}^l z_j$,
where $l=1, \dots N$.

We are mainly interested in the description of the differential distributions for events with fixed final state multiplicity, firstly on a qualitative level. At this point we need to note the following:
the final jet multiplicity for a given final state is uniquely defined, it actually depends on the lower $p_T$ cutoff
we set for a mini-jet to qualify as a jet as well as on the chosen jet radius $R$ (in the rapidity-azimuthal angle plane) for the jet clustering algorithm. Nevertheless, we believe that our analysis in the following is valid
 once a well defined mechanism for deciding the multiplicity of a final state is established. 
This is not a trivial statement at is it implies that if a final state has initially been assigned  multiplicity $N_1+2$ complies with the $N_1+2$ differential distributions, then, if a different set of parameters is chosen for the jet clustering algorithm which results in a different number of final state jets giving a shift
in multiplicity from $N_1$ to $N_2$,  then that final state will also comply with the $N_2+2$ differential distributions. 

\begin{figure}
\begin{center}
\begin{flushleft}
\hspace{1.cm}\includegraphics[width=7cm]{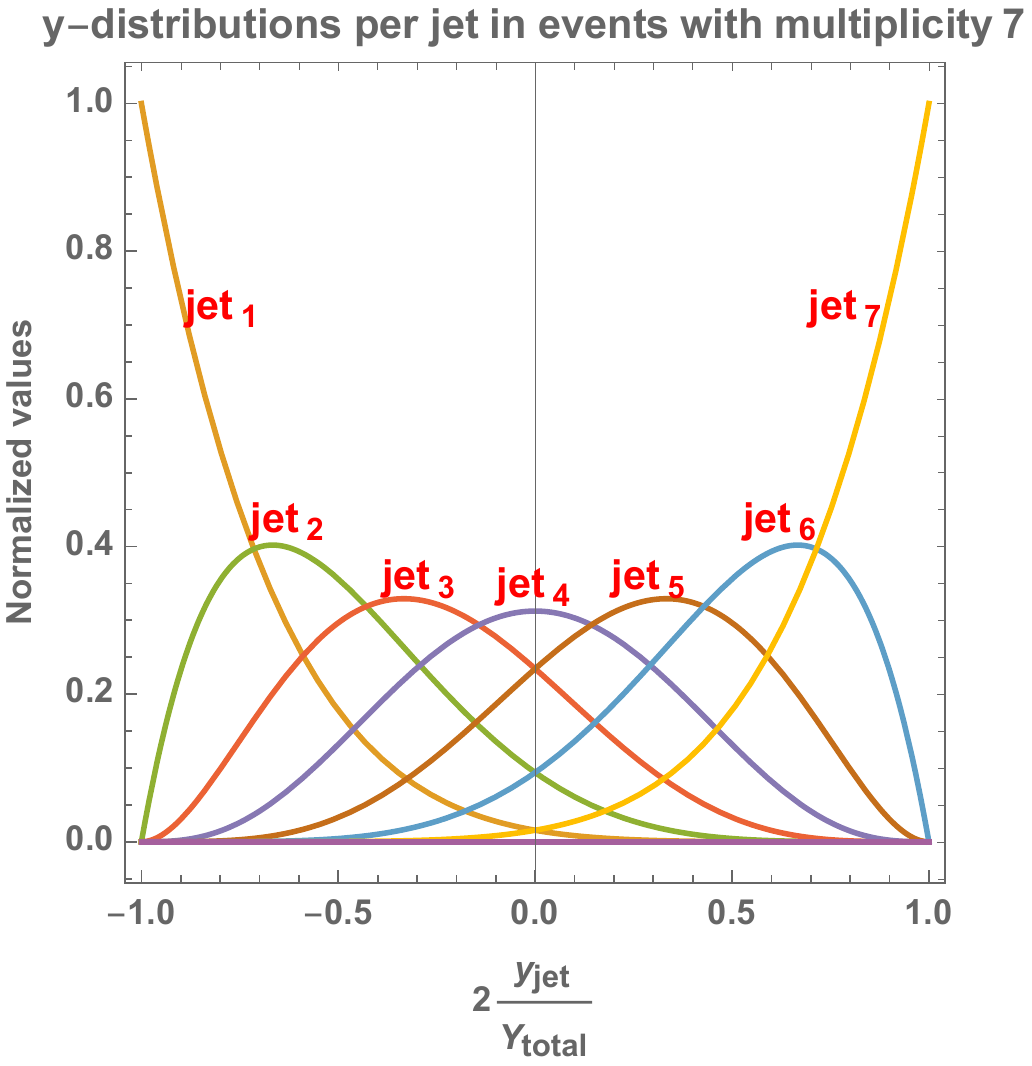}
\end{flushleft}
\vspace{-7.6cm}
\begin{center}
\hspace{8cm}\includegraphics[width=7.cm]{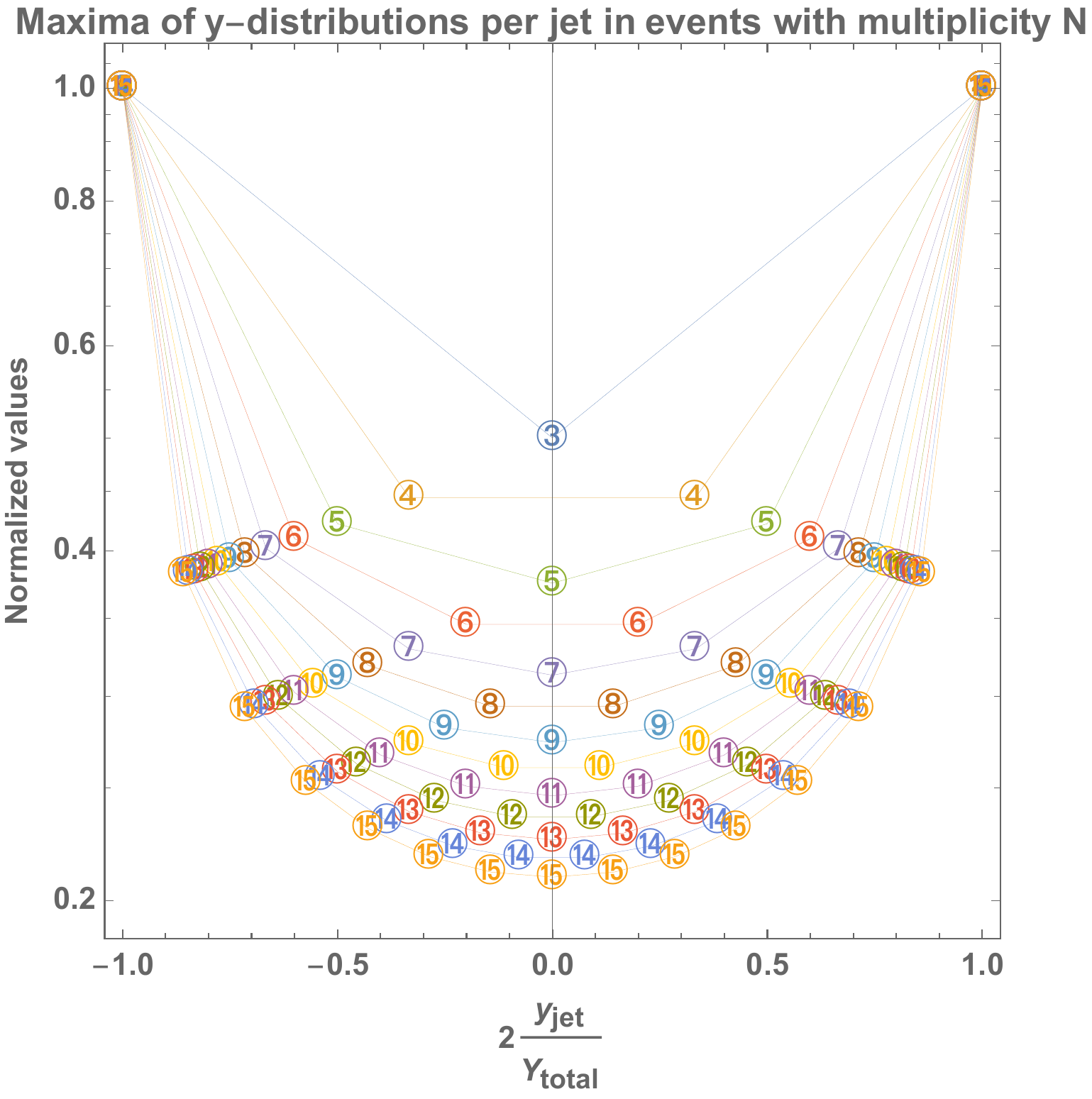}
\end{center}
\vspace{-.5cm}
\caption{Rapidity distributions for each of the jets in a final state with seven mini-jets (left).  Their maxima are indicated by the symbol \textcircled{\tiny 7} (right). The positions of the $y$-distribution maxima in configurations with multiplicity $N$ are marked by \textcircled{\tiny N} (right).}
\label{Multiplicity7yjets}
\end{center}
\end{figure}

The contribution for jet $l$ in a $N+2$ final state event, will have the following differential distribution in rapidity 
\begin{eqnarray}
\frac{d \sigma_{N+2}^{(l)}}{d y_l} &=&  \alpha^{N+2} \int_0^Y 
  \prod_{i=1}^{N+1} dz_i \delta \left(Y-\sum_{s=1}^{N+1} z_s \right) 
\delta \left(y_l +\frac{Y}{2}- \sum_{j=1}^l z_j\right) \nonumber\\
&=& \alpha^{N+2} \int_{y_l}^\frac{Y}{2} dy_N \int_{y_l}^{y_N} dy_{N-1} \cdots \int_{y_l}^{y_{l+2}} dy_{l+1}
\int_{-\frac{Y}{2}}^{y_l} dy_{l-1}  \cdots \int_{-\frac{Y}{2}}^{y_3} dy_2 
\int_{-\frac{Y}{2}}^{y_2} dy_1\nonumber\\
&=&  \alpha^{N+2} \frac{\left(\frac{Y}{2}-y_l \right)^{N-l}}{(N-l)!} \frac{\left(y_l+\frac{Y}{2}\right)^{l-1}}{(l-1)!}  \, ,
\label{dsdy}
\end{eqnarray}
as derived from Eq.~(\ref{sigma_N2}).
For very large multiplicities this results to an asymptotic Poisson distribution as one can verify, {\it e.g.}, in the region $y \simeq - \frac{Y}{2}$ with $y = \left(\frac{\lambda}{N}-\frac{1}{2}\right) Y$ where 
\begin{eqnarray}
\lim_{N \to \infty}{N-1 \choose l-1}   
\left(1-\frac{\lambda}{N}\right)^{N-l} 
\left(\frac{\lambda}{N}\right)^{l-1}
&=& e^{- \lambda} \frac{\lambda^{l-1}}{(l-1)!} \, .
\end{eqnarray}
Taking the limits $l \to 1$ and $y_l \to - \frac{Y}{2}$ in 
Eq.~(\ref{dsdy}) we derive a normalized universal distribution for each $N$ when plotted versus $2y/Y$. The case for  N=7+2 is plotted in Fig.~\ref{Multiplicity7yjets} (left) which is very characteristic for multiperipheral models. We remind the reader that the notation jet$_{i=1,2, \dots, N}$ is introduced for jets with ordered rapidities $y_1 < y_2 < \dots < y_N$.
All these seven normalized $y$-distributions (one for each of the seven jets) spans an area of $\frac{2}{N}$. Their maxima are found  at $y=\frac{2l-N-1}{2(N-1)} Y$ with a value that is
\begin{eqnarray}
{N-1 \choose l-1} \frac{(l-1)^{l-1}}{(N-1)^{N-1}} \left(N- l\right)^{N-l} \, .
\end{eqnarray}
In Fig.~\ref{Multiplicity7yjets} (right), we show the maxima for multiplicities  $N$, where $\left(3+2\right) \le \left(N+2\right) \le \left(15+2\right)$.

In a similar manner, we get expressions for the double differential rapidity distributions for jet pairs, {\it i.e.}
\begin{eqnarray}
\frac{d^2 \sigma_{N+2}^{(l,m)}}{d y_l d y_m} &=&  \alpha^{N+2} 
\int_{0}^{Y}
  \prod_{i=1}^{N+1} dz_i \delta \left(Y-\sum_{s=1}^{N+1} z_s \right) 
\delta \left(y_l +\frac{Y}{2}- \sum_{j=1}^l z_j\right)
\delta \left(y_m +\frac{Y}{2}- \sum_{k=1}^m z_k\right) \nonumber\\
&=&  \alpha^{N+2} \frac{\left(\frac{Y}{2}-y_l \right)^{N-l}}{(N-l)!}
\frac{(y_l-y_m)^{l-m-1}}{(l-m-1)!} 
\frac{\left(y_m+\frac{Y}{2}\right)^{m-1}}{(m-1)!}  \, .
\label{d2sdydy}
\end{eqnarray}

To calculate the correlation between the rapidities of jet $l$ and jet $m$ we use Eq.~(\ref{correlation_function}),
more precisely 
\begin{eqnarray}
{\cal R}_{N+2} \left(x_l,x_m\right) = \sigma_{N+2} 
\frac{ \frac{ d^2 \sigma_{N+2}^{(l,m)}}{d y_l d y_m} }{\frac{d \sigma_{N+2}^{(l)}}{d y_l} \frac{d \sigma_{N+2}^{(m)}}{d y_m}}-1 
 =  
\frac{2^N}{N!}\frac{(N-m)!(l-1)!}{(l-m-1)!}   
\frac{(x_l-x_m)^{l-m-1}}{\left(1+x_l\right)^{l-1}\left(1-x_m \right)^{N-m}}
 -1
\, ,
\label{R}
\end{eqnarray}
where $Y > y_l > y_m > 0$, $l>m$ and $x_J = 2 y_J / Y$.

\begin{figure}
\begin{subfigure}{.5\textwidth}
  \centering
  \includegraphics[width=1\linewidth]{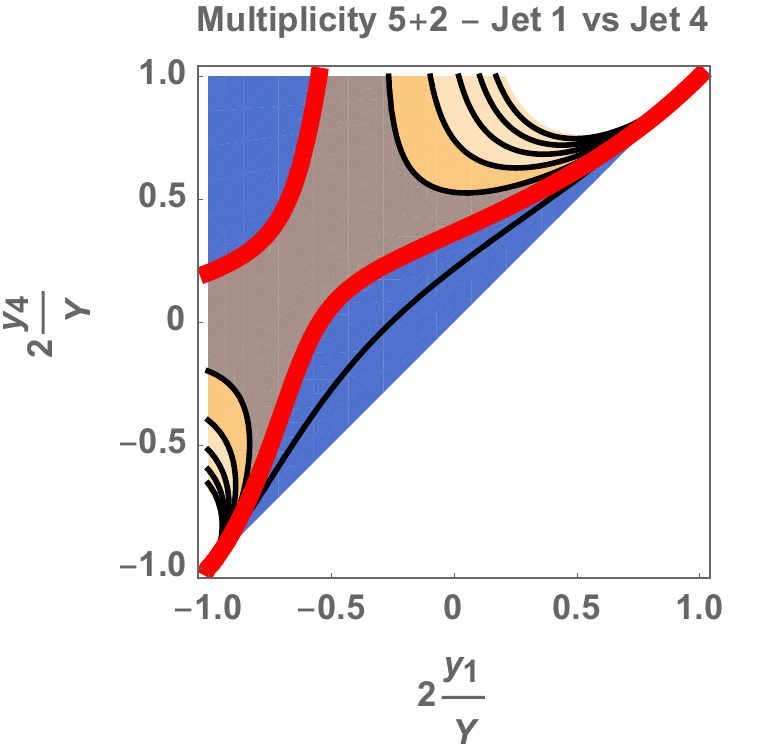}
  \label{fig:sfig1}
\end{subfigure}%
\begin{subfigure}{.5\textwidth}
  \centering
  \includegraphics[width=1\linewidth]{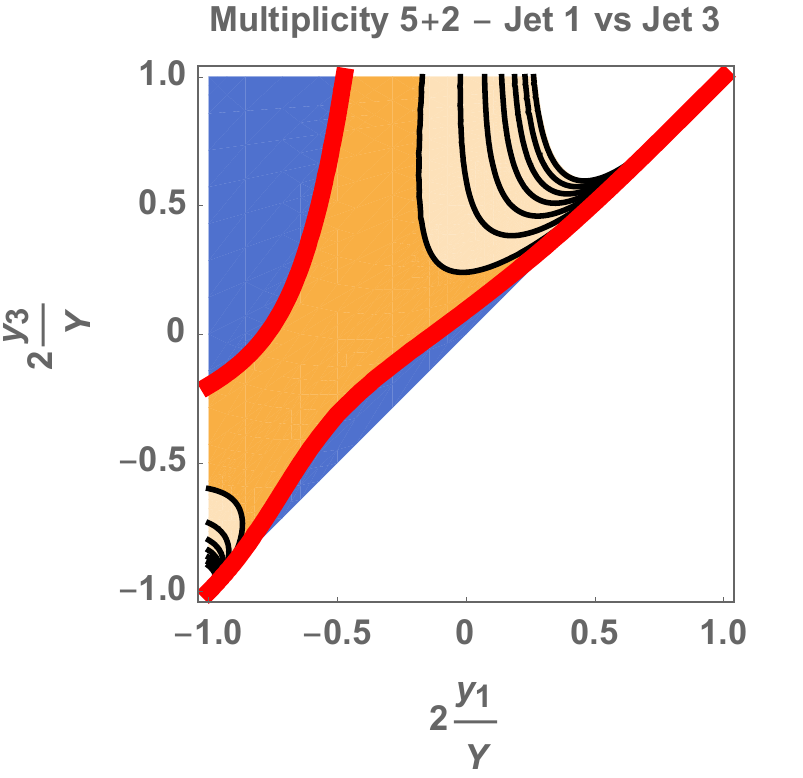}
  \label{fig:sfig2}
\end{subfigure}
\caption{Left:  ${\cal R}_{5+2} \left(x_4,x_1\right) = \sigma_{5+2} 
\frac{ \frac{ d^2 \sigma_{5+2}^{(4,1)}}{d y_4 d y_1} }{\frac{d \sigma_{5+2}^{(4)}}{d y_4} \frac{d \sigma_{5+2}^{(1)}}{d y_1}}-1$. 
Right:  ${\cal R}_{5+2} \left(x_3,x_1\right) = \sigma_{5+2} 
\frac{ \frac{ d^2 \sigma_{5+2}^{(3,1)}}{d y_3 d y_1} }{\frac{d \sigma_{5+2}^{(3)}}{d y_3} \frac{d \sigma_{5+2}^{(1)}}{d y_1}}-1$. }
\label{R7-1425}
\end{figure}

\begin{figure}
\begin{subfigure}{.5\textwidth}
  \centering
  \includegraphics[width=1\linewidth]{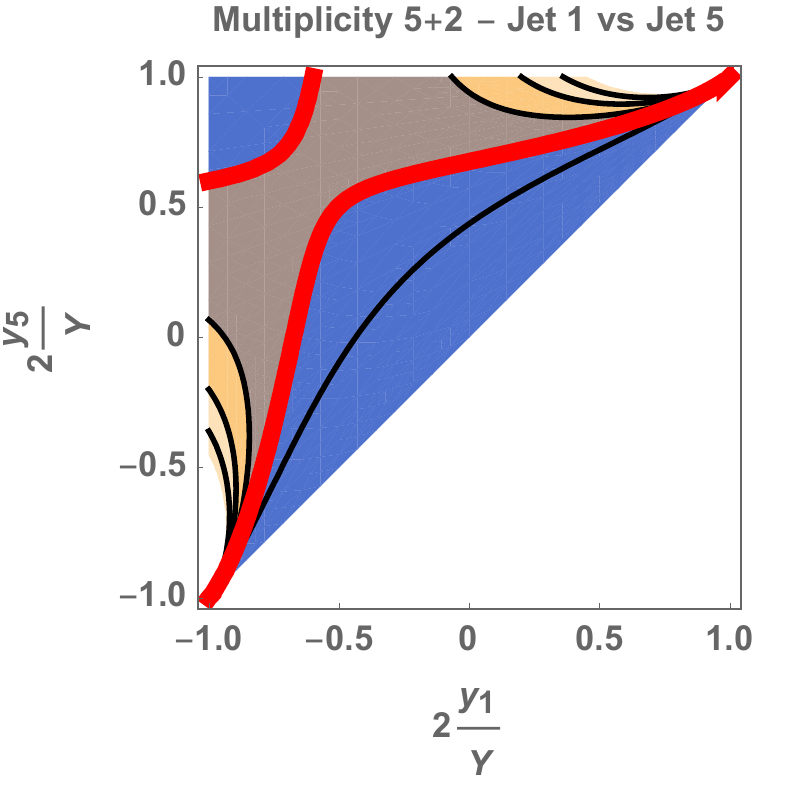}
  \label{fig:sfig1}
\end{subfigure}%
\begin{subfigure}{.5\textwidth}
  \centering
  \includegraphics[width=1\linewidth]{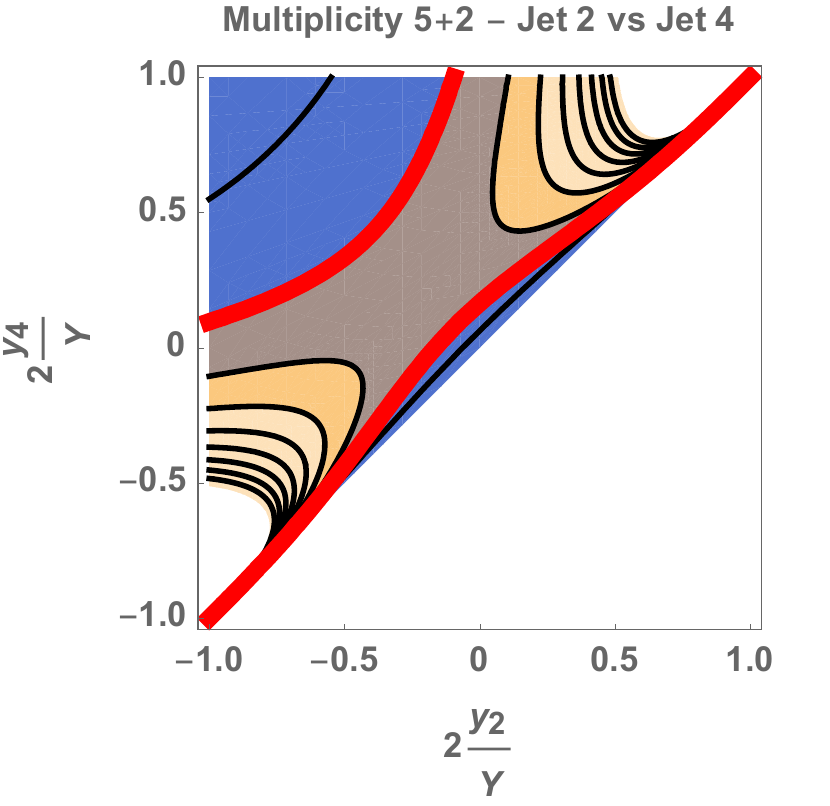}
  \label{fig:sfig2}
\end{subfigure}
\caption{Left:  ${\cal R}_{5+2} \left(x_5,x_1\right) = \sigma_{5+2} 
\frac{ \frac{ d^2 \sigma_{5+2}^{(5,1)}}{d y_5 d y_1} }{\frac{d \sigma_{5+2}^{(5)}}{d y_5} \frac{d \sigma_{5+2}^{(1)}}{d y_1}}-1$. 
Right:  ${\cal R}_{5+2} \left(x_4,x_2\right) = \sigma_{5+2} 
\frac{ \frac{ d^2 \sigma_{5+2}^{(4,2)}}{d y_4 d y_2} }{\frac{d \sigma_{5+2}^{(4)}}{d y_4} \frac{d \sigma_{5+2}^{(2)}}{d y_2}}-1$. }
\label{R7-1524}
\end{figure}

We plot the correlation functions using Eq.~(\ref{R}) in Figs.~\ref{R7-1425} and ~\ref{R7-1524}. We choose
multiplicity $5+2$ and we show the correlation between the rapidities of jet 1 and jet 4 (Fig.~\ref{R7-1425}, left),
jet 1 and jet 3 (Fig.~\ref{R7-1425}, right), jet 1 and jet 5 (Fig.~\ref{R7-1524}, left), jet 2 and jet 4 (Fig.~\ref{R7-1524}, right). The red lines in each of them underine the contour for which ${\cal R}=0$ while the white regions correspond to sectors of very rapid growth of ${\cal R}$. 
\begin{figure}
\begin{subfigure}{.5\textwidth}
  \centering
  \includegraphics[width=.7\linewidth]{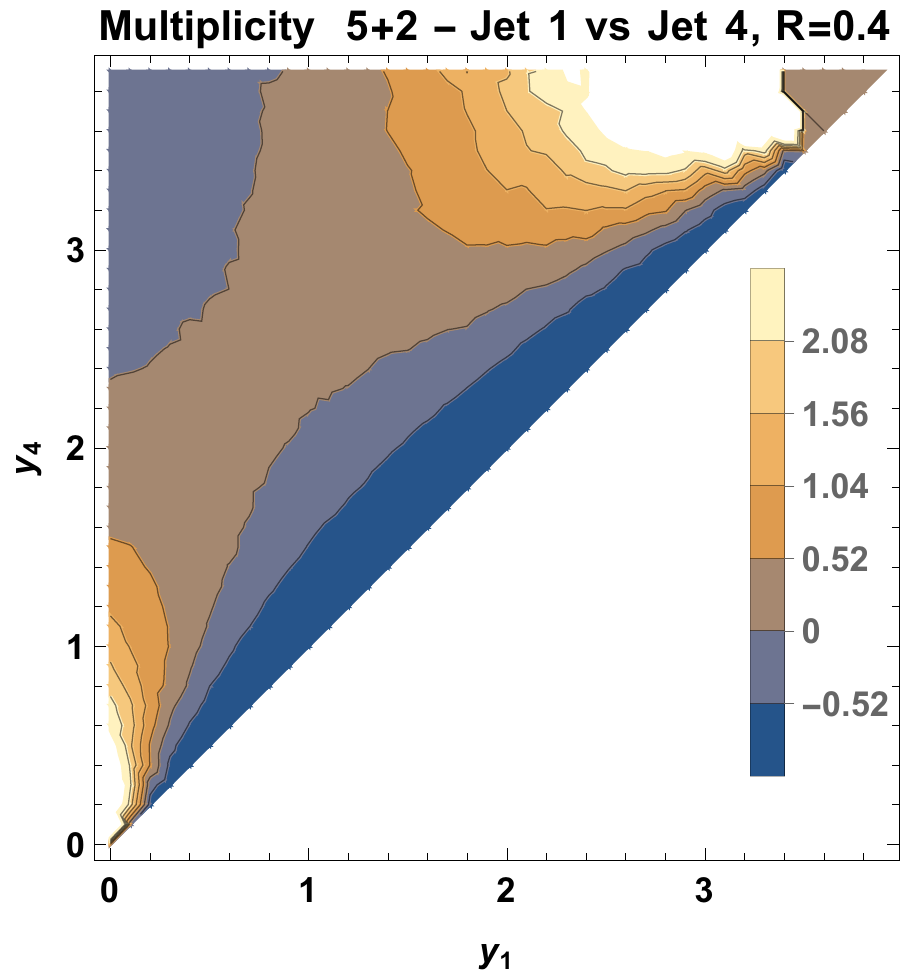}
  \caption{}
  \label{fig:sfig1}
\end{subfigure}%
\begin{subfigure}{.5\textwidth}
  \centering
  \includegraphics[width=.7\linewidth]{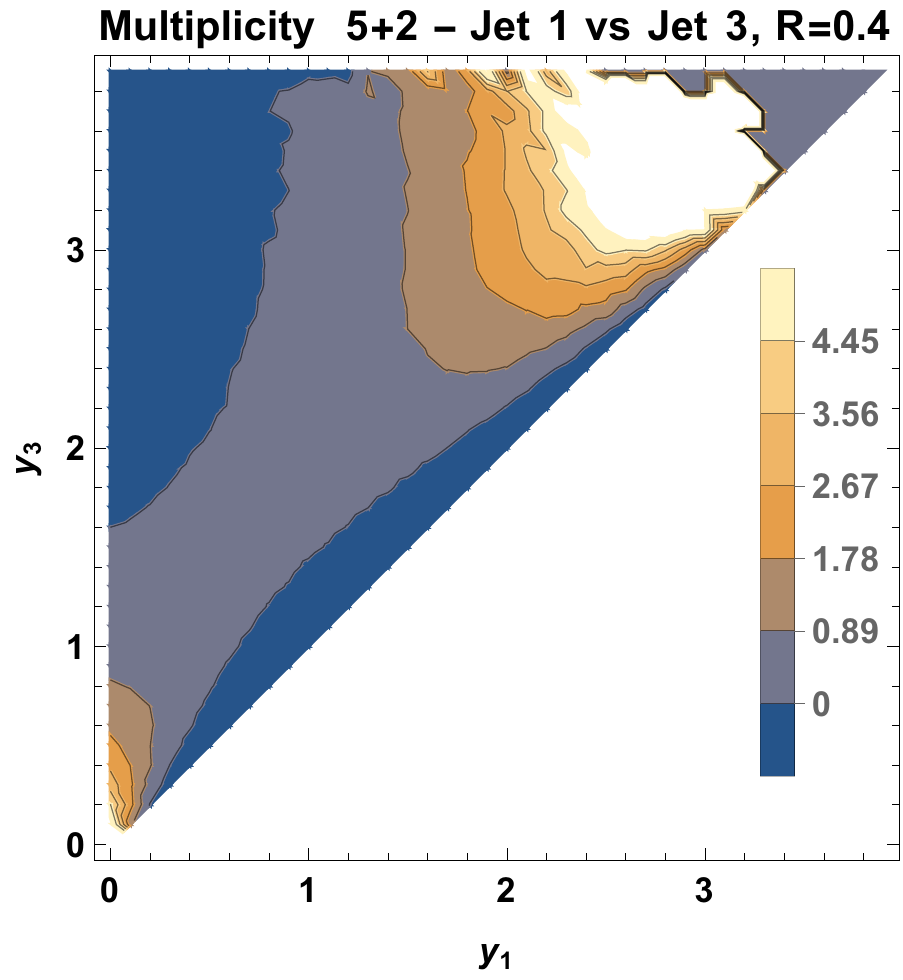}
  \caption{}
  \label{fig:sfig2}
\end{subfigure}
\\
\begin{subfigure}{.5\textwidth}
  \centering
  \includegraphics[width=.7\linewidth]{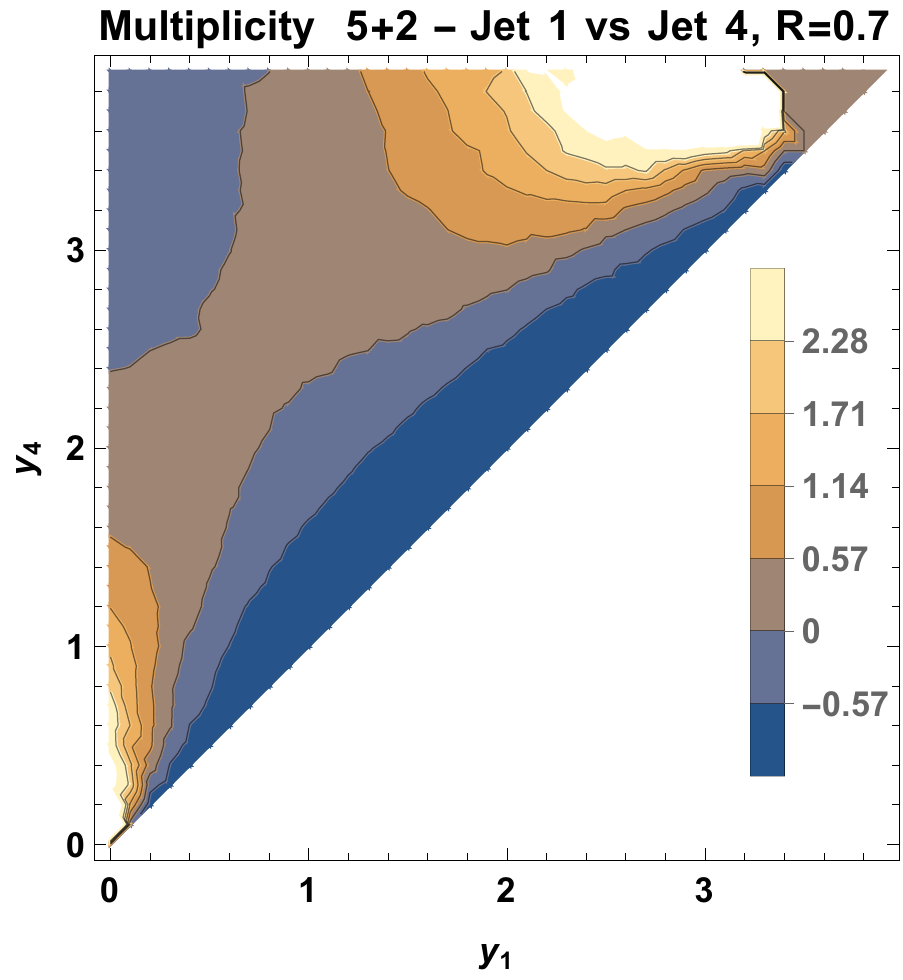}
  \caption{}
  \label{fig:sfig1}
\end{subfigure}%
\begin{subfigure}{.5\textwidth}
  \centering
  \includegraphics[width=.7\linewidth]{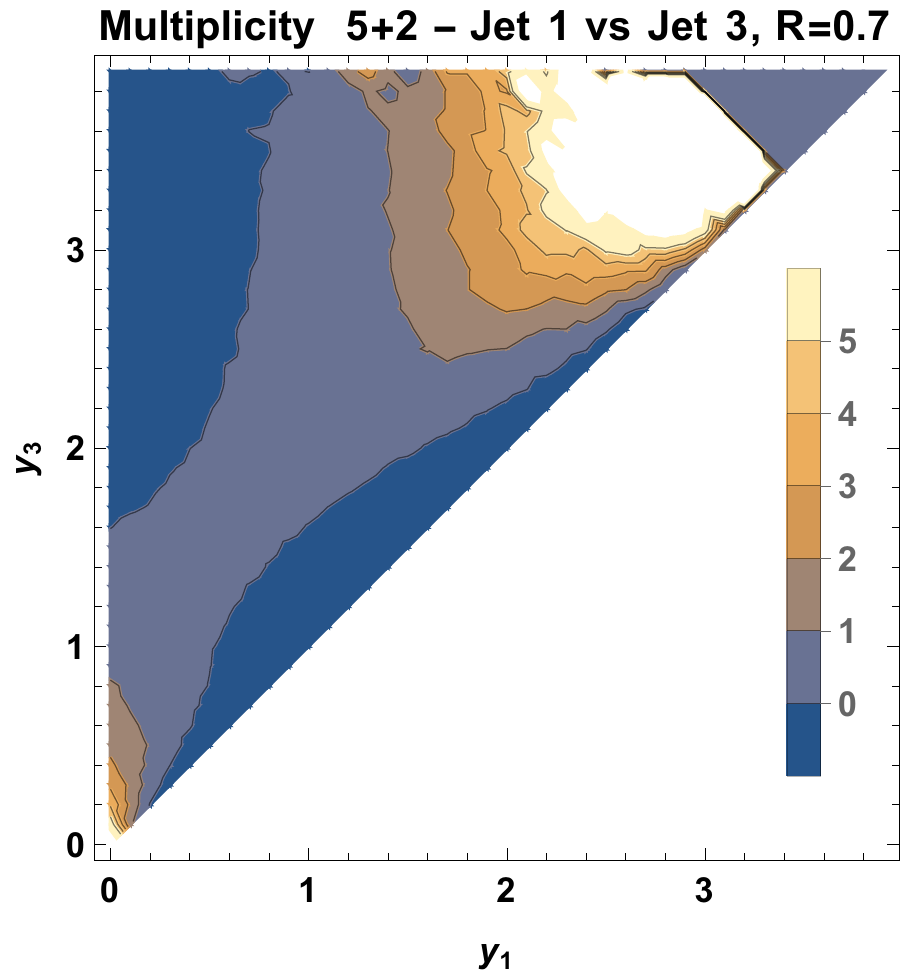}
  \caption{}
  \label{fig:sfig2}
\end{subfigure}
\caption{Top: The correlation functions of Fig.~\ref{R7-1425} with the collinear BFKL model and for jet radius $R = 0.4.$
Bottom: The same but for jet radius $R = 0.7$.}
\label{fig:coll1}
\end{figure}

\begin{figure}
\begin{subfigure}{.5\textwidth}
  \centering
  \includegraphics[width=.7\linewidth]{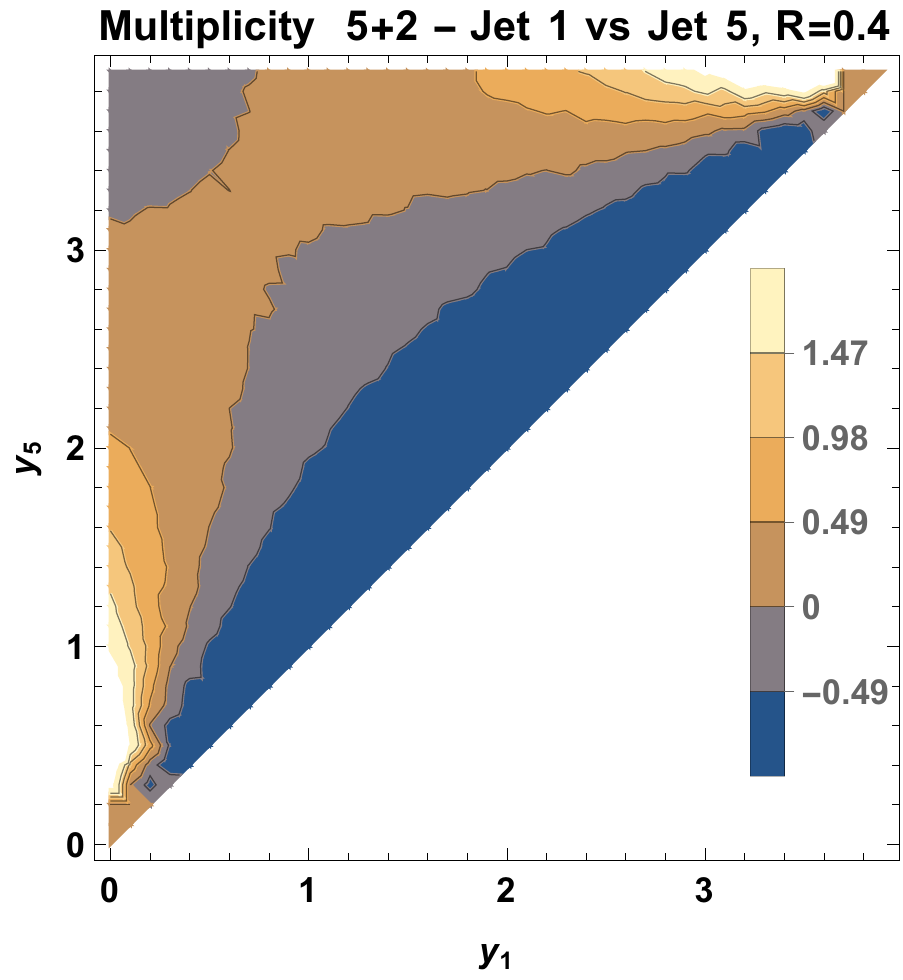}
  \caption{}
  \label{fig:sfig1}
\end{subfigure}%
\begin{subfigure}{.5\textwidth}
  \centering
  \includegraphics[width=.7\linewidth]{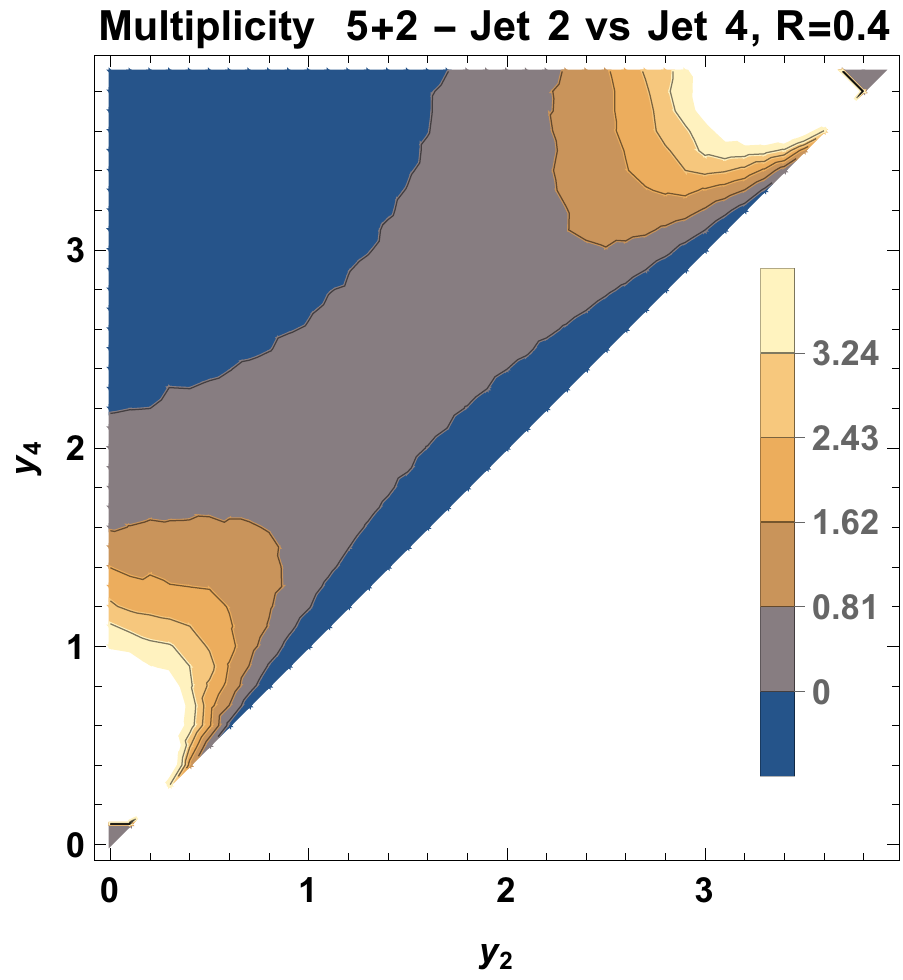}
  \caption{}
  \label{fig:sfig2}
\end{subfigure}
\\
\begin{subfigure}{.5\textwidth}
  \centering
  \includegraphics[width=.7\linewidth]{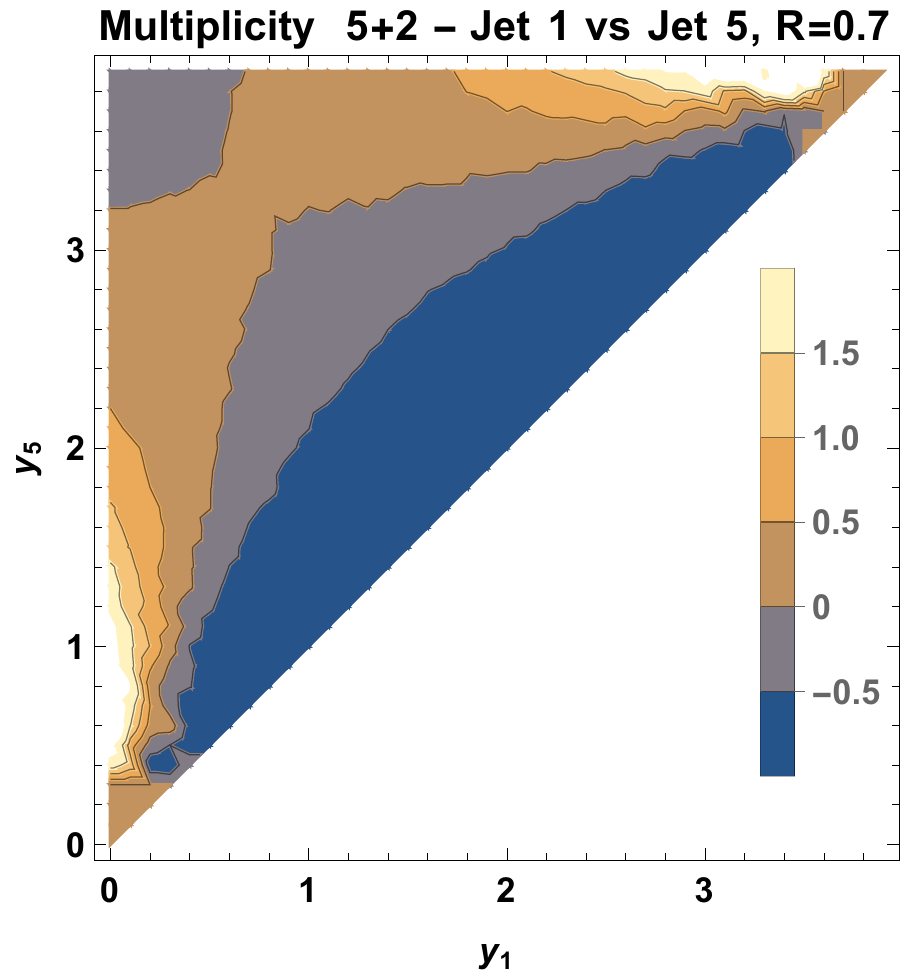}
  \caption{}
  \label{fig:sfig1}
\end{subfigure}%
\begin{subfigure}{.5\textwidth}
  \centering
  \includegraphics[width=.7\linewidth]{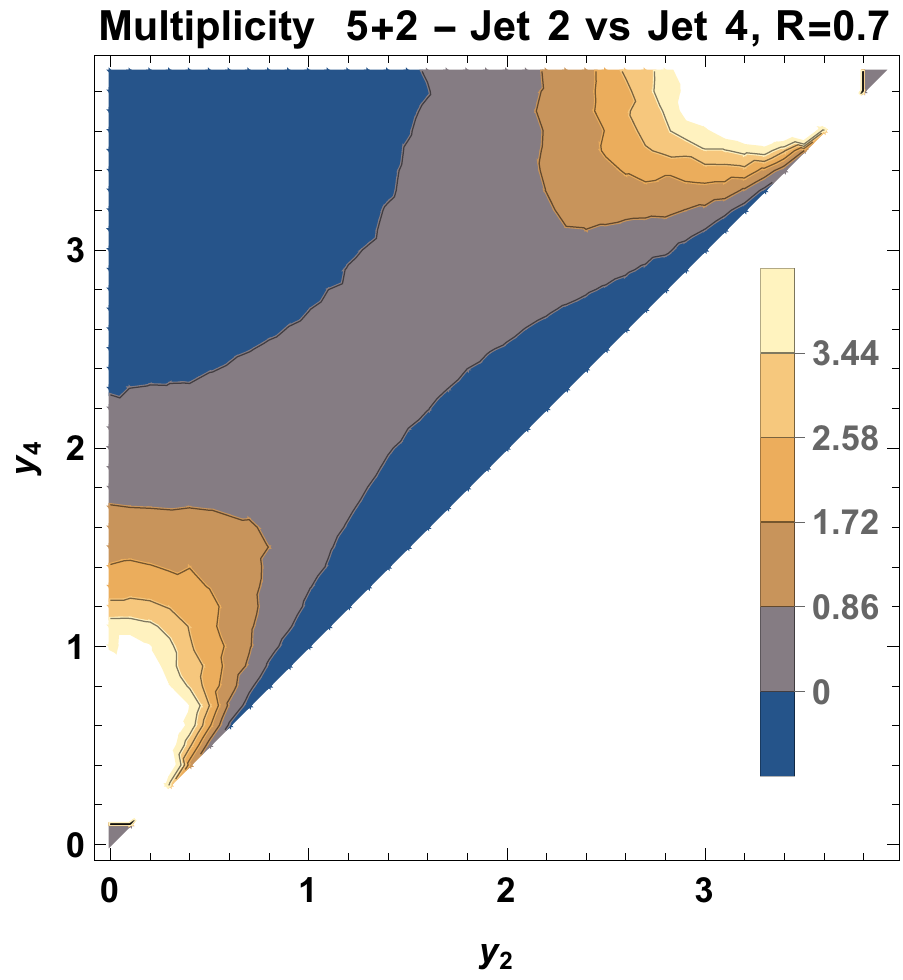}
  \caption{}
  \label{fig:sfig2}
\end{subfigure}
\caption{Top: The correlation functions of Fig.~\ref{R7-1524} with the collinear BFKL model and for jet radius $R = 0.4.$
Bottom: The same but for jet radius $R = 0.7$.}
\label{fig:coll2}
\end{figure}

Our analysis here is quite simple as no relevant dynamics from the  transverse coordinates of the phase space is introduced. Nevertheless, it should still be enough to capture the gross features of single rapidity distributions and double rapidity correlations which should be generally independent of the selected $p_T$ of the outermost
in rapidity jets. In the next section, we will investigate whether these gross
features are any similar to predictions derived from the BFKL formalism using a simple collinear BFKL model
implemented as a special case in our Monte Carlo code {\tt BFKLex}~\cite{Chachamis:2011rw,Chachamis:2011nz,Chachamis:2012fk,Chachamis:2012qw,Chachamis:2015zzp,Chachamis:2015ico,deLeon:2021ecb}.
 
\section{Correlations in BFKL with BFKLex}
 
 In this section we evaluate the correlation functions for the rapidities of jets within the BFKL formalism. We 
 work with the gluon Green's function, that is at partonic level, where an emitted gluon is named a jet if its
 $p_T$ is above some cutoff and we leave for a future work the 
 more complicated and detailed study of the full BFKL dynamics at hadronic level.
 
At a collider with colliding energy $s$, at the leading logarithmic approximation,
where large logarithmic terms of the form ${\bar \alpha}_s^n \ln^n{s}$ are resummed 
(using ${\bar \alpha}_s= {\bar \alpha}_s N_c  / \pi$), 
the differential partonic cross section for the production of two well separated in
rapidity jets and transverse momenta $\vec{p}_{i=1,2}$ is given by
 \begin{eqnarray}
\frac{d \hat{\sigma}}{d^2 \vec{p}_1 d^2 \vec{p}_2} &=& 
\frac{\pi^2 {\bar \alpha}_s^2}{2} \frac{f(\vec{p}_1^{~2}, \vec{p}_2^{~2},Y)}{\vec{p}_1^{~2} \vec{p}_2^{~2}} \,.
\end{eqnarray}
where we take the longitudinal momentum fractions of the colliding partons to be $x_{i=1,2}$ 
and the rapidity difference between the two jets $Y \sim \ln{x_1 x_2 s / \sqrt{\vec{p}_1^{~2} \vec{p}_2^{~2}}}$ .

Within the BFKL resummation framework, one can show that the gluon Green's function $f$ follows, in a collinear approximation, the integro-differential equation
\begin{eqnarray}
\frac{\partial f (K^2,Q^2,Y) }{\partial Y} &=& \delta (K^2-Q^2) \nonumber \\
&&\hspace{-1cm}+ \, 
 {\bar \alpha}_s \int_0^\infty d q^2 \left(\frac{\theta (K-q)}{K^2}+\frac{\theta (q-K)}{q^2} + 4 (\ln{2}-1) \delta(q^2-K^2)\right)
f (q^2,Q^2,Y) \, ,
\end{eqnarray}
the solution of which can be cast in an iterative form:
\begin{eqnarray}
f (K^2,Q^2,Y) &=& e^{4(\ln{2}-1){\bar \alpha}_s Y} \Bigg\{\delta (K^2-Q^2)    \nonumber\\
&&\hspace{-1cm}+ \,    \sum_{N=1}^\infty 
 \frac{({\bar \alpha}_s Y)^N}{N!} \left[ \prod_{L=1}^N
\int_0^\infty d x_L \left(\frac{\theta(x_{L-1}-x_L)}{x_{L-1}} + \frac{\theta(x_L-x_{L-1})}{x_{L}}\right) \right]
\delta (x_N-Q^2) \Bigg\} \,.
\label{FCSumMC}
\end{eqnarray}
Using 
$\delta(K^2-Q^2)=\int \frac{d \gamma}{2 \pi i Q^2} \left(\frac{K^2}{Q^2}\right)^{\gamma-1}$ 
and 
\begin{eqnarray}
&&\int_0^\infty d x_N 
\left(
\frac{\theta(x_{N-1}-x_N)}{x_{N-1}} 
+ \frac{\theta(x_N-x_{N-1})}{x_{N}}\right) 
\left(\frac{x_N}{K^2}\right)^{\gamma-1} = \left(\frac{1}{\gamma}+\frac{1}{1-\gamma}\right) \left(\frac{x_{N-1}}{K^2}\right)^{\gamma-1},
\end{eqnarray}
which is valid for $0<\gamma<1$, we can write
\begin{eqnarray}
\left[ \prod_{L=1}^N \int_0^\infty d x_L 
\left(\frac{\theta(x_{L-1}-x_L)}{x_{L-1}} 
+ \frac{\theta(x_L - x_{L-1})}{x_{L}}\right) \right]
\left(\frac{x_N}{K^2}\right)^{\gamma-1}  
= \left(\frac{1}{\gamma}+\frac{1}{1-\gamma}\right)^N
\end{eqnarray}
so that we then have
\begin{eqnarray}
f (K^2,Q^2,Y) &=&  \frac{e^{4(\ln{2}-1) {\bar \alpha}_s Y}}{Q^2} \int \frac{d \gamma}{2 \pi i} \left(\frac{K^2}{Q^2}\right)^{\gamma-1}
 \sum_{N=0}^\infty \frac{\left({\bar \alpha}_s Y\right)^N}{N!}
\left(\frac{1}{\gamma}+\frac{1}{1-\gamma}\right)^N  \nonumber\\
&=&  \int \frac{d \gamma}{2 \pi i Q^2} \left(\frac{K^2}{Q^2}\right)^{\gamma-1} e^{{\bar \alpha}_s Y \chi\left( \gamma\right)} \, ,
\label{FCMM}
\end{eqnarray}
with $\chi(\gamma) = 4 (\ln{2}-1)+ \gamma^{-1}+ (1-\gamma)^{-1}$. 

From a simple inspection of Eq.~\ref{FCSumMC} one can infer that for the class of jet-jet rapidity correlations we studied in the Section 2, the predictions from this collinear BFKL model
should be very similar to those in the Chew-Pignotti simple model. 
Actually, in Eq.~(\ref{FCSumMC}) every $N$ increase by one unit
accounts for the emission of a new final state gluon.  
After making use of Eq.~(\ref{sigma_N2}), we can write
\begin{eqnarray}
f (K^2,Q^2,Y) &=&  \sum_{N=0}^\infty  {\bar \alpha}_s^{N} 
\int_{0}^{Y} \prod_{i=1}^{N+1} dz_i \delta \left(Y-
\sum_{s=1}^{N+1} z_s \right) \xi^{(N)} (K^2,Q^2) 
\end{eqnarray}
where
\begin{eqnarray}
\xi^{(N)} (K^2,Q^2) &=& \int \frac{d \gamma}{2 \pi i Q^2} \left(\frac{K^2}{Q^2}\right)^{\gamma-1}  \chi^N\left( \gamma\right) \, .
\end{eqnarray}
Working in a manner that follows the logic we used to obtain Eqs.~(\ref{dsdy}) and 
(\ref{d2sdydy}), {\it i.e.} we get here
\begin{eqnarray}
\frac{d f_{N}^{(l)} (K^2,Q^2,Y, y_l) }{d y_l} &=&  {\bar \alpha}_s^{N} \frac{\left(\frac{Y}{2}-y_l \right)^{N-l}}{(N-l)!} \frac{\left(y_l+\frac{Y}{2}\right)^{l-1}}{(l-1)!} \xi^{(N)} (K^2,Q^2)  \, ,  \\
 \frac{d f_{N}^{(l,m)} (K^2,Q^2,Y, y_l,y_m) }{d y_l d y_m} &=& 
 {\bar \alpha}_s^{N} \frac{\left(\frac{Y}{2}-y_l \right)^{N-l}}{(N-l)!}
\frac{(y_l-y_m)^{l-m-1}}{(l-m-1)!} 
\frac{\left(y_m+\frac{Y}{2}\right)^{m-1}}{(m-1)!} \xi^{(N)} (K^2,Q^2)   \, .
\end{eqnarray}
Obviously,  the 
$\xi^{(N)}$ factor cancels out  for normalized quantities
 and thus we end up with the same expressions as for 
the Chew-Pignotti model. 
It is important to remember at this point that the full BFKL formalism 
carries non-trivial dependences in rapidity, transverse momenta and azimuthal angles 
which need to be studied in detail in future works. 
Nevertheless, our findings here in rapidity space suggest that the 
full BFKL predictions might not be totally different from
the old multiperipheral model approach. 

To connect with future works, we have implemented the BFKL collinear model within our {\tt BFKLex} Monte Carlo code, setting the transverse momenta of the most forward and backward jets to be 30 and 35 GeV 
respectively and their difference in rapidity $Y=4$. We also set the multiplicity of the emitted gluons (jets) to
be $N=5+2$. We use the anti-kt jet clustering algorithm in its implementation
 in {\tt fastjet}~\cite{Cacciari:2011ma}. 
We present results from the collinear model  in Figs.~\ref{fig:coll1} and~\ref{fig:coll2}.
The jet radius was chosen to take two values, $R = 0.4$ and $R = 0.7$. In Fig.~\ref{fig:coll1} we plot
the same correlation functions as in Fig.~\ref{R7-1425} whereas in Fig.~\ref{fig:coll2}
the corresponding ones for Fig.~\ref{R7-1524}. It is not surprising, that the collinear model results are 
very similar to the Chew-Pignotti ones and obviously the actual jet radius R does not
affect significantly the plots. Let us also note that in Figs.~\ref{fig:coll1} and~\ref{fig:coll2}
we kept the rapidity range from 0 to $Y$  to make the association to experimental data setups easier. Such
can be found for example in relevant dijet experimental analyses for 7 TeV data from both ATLAS and CMS~\cite{Aad:2011jz,Chatrchyan:2012pb,Aad:2014pua,Khachatryan:2016udy}.

\section{Conclusions}
The 13 TeV data from the run 2 of the LHC at low luminosity are suitable for
various studies of multi-jet physics. Following our recent work in Ref.~\cite{deLeon:2020myv}, 
we want to suggest the investigation of a particular subset of Mueller-Navelet jet events where the outermost jets
are very similar in $p_T$ and the jet multiplicity is kept fixed.
We believe that their experimental study is interesting as it might be possible to  identify features of different 
multi-particle production models such as those predicted by the BFKL formalism. 
We have presented predictions for single and double differential distributions in jet rapidity as well as
the jet-jet correlation functions
from an old multiperipheral model, namely the Chew-Pignotti model, using analytic expressions we obtained
after performing an analysis based on the decoupling of the longitudinal and transverse coordinates. 
We have also presented results for the jet-jet correlation
functions from a collinear BFKL model implemented in our Monte Carlo code {\tt BFKLex}.
In the future, we plan to perform a more complete study of these observables in high energy QCD
including the full dependence on the transverse coordinates
and moving from a partonic level analysis of the BFKL gluon Green's function 
to the hadronic level with PDFs included and suitable phenomenological kinematic cuts.

Comments: Presented at the Low-$x$ Workshop, Elba Island, Italy, September 27--October 1 2021.

\section*{Acknowledgements}
We would like to thank the organizers of the Low-x Workshop for their excellent work. 
This work has been supported by the Spanish Research Agency (Agencia Estatal de Investigaci{\'o}n) through the grant IFT Centro de Excelencia Severo Ochoa SEV-2016-0597 and the Spanish Government grant FPA2016-78022-P.  It has also received funding from the European Union's Horizon 2020 research
and innovation programme under grant agreement No. 824093. The work of GC was supported by the Funda\c{c}{\~ a}o para a Ci{\^ e}ncia e a Tecnologia (Portugal) under project CERN/FIS-PAR/0024/2019 and contract 'Investigador auxiliar FCT - Individual Call/03216/2017'.


\begin{thebibliography}{}

%%%%%Dremin
%\cite{Dremin:1977wc}
\bibitem{Dremin:1977wc}
I.~M.~Dremin, C.~Quigg,
%``The Cluster Concept in Multiple Hadron Production,''
Science \textbf{199}, 937 (1978). 
%doi:10.1126/science.199.4332.937
%14 citations counted in INSPIRE as of 16 Dec 2020

%\cite{Tannenbaum:2005by}
\bibitem{Tannenbaum:2005by}
M.~J.~Tannenbaum,
%``From the ISR to RHIC: Measurements of hard-scattering and jets using inclusive single particle production and 2-particle correlations,''
J. Phys. Conf. Ser. \textbf{27}, 1-10 (2005).
%doi:10.1088/1742-6596/27/1/001
%[arXiv:nucl-ex/0507020 [nucl-ex]].
%10 citations counted in INSPIRE as of 17 Dec 2020


%\cite{SanchisLozano:2008te}
\bibitem{SanchisLozano:2008te}
M.~A.~Sanchis-Lozano,
%``Prospects of searching for (un)particles from Hidden Sectors using rapidity correlations in multiparticle production at the LHC,''
Int. J. Mod. Phys. A \textbf{24}, 4529-4572 (2009). 
%doi:10.1142/S0217751X09045820
%[arXiv:0812.2397 [hep-ph]].
%16 citations counted in INSPIRE as of 17 Dec 2020


%%%%% DGLAP
%\cite{Gribov:1972ri}
\bibitem{Gribov:1972ri}
V.~N.~Gribov, L.~N.~Lipatov,
%``Deep inelastic e p scattering in perturbation theory,''
Sov. J. Nucl. Phys. \textbf{15}, 438-450 (1972)
IPTI-381-71.
%4506 citations counted in INSPIRE as of 16 Dec 2020

%\cite{Gribov:1972rt}
\bibitem{Gribov:1972rt}
V.~N.~Gribov, L.~N.~Lipatov,
%``e+ e- pair annihilation and deep inelastic e p scattering in perturbation theory,''
Sov. J. Nucl. Phys. \textbf{15}, 675-684 (1972).
%1327 citations counted in INSPIRE as of 16 Dec 2020

%\cite{Lipatov:1974qm}
\bibitem{Lipatov:1974qm}
L.~N.~Lipatov,
%``The parton model and perturbation theory,''
Sov. J. Nucl. Phys. \textbf{20}, 94-102 (1975).
%1556 citations counted in INSPIRE as of 16 Dec 2020

%\cite{Altarelli:1977zs}
\bibitem{Altarelli:1977zs}
G.~Altarelli, G.~Parisi,
%``Asymptotic Freedom in Parton Language,''
Nucl. Phys. B \textbf{126}, 298-318 (1977).
%doi:10.1016/0550-3213(77)90384-4
%7280 citations counted in INSPIRE as of 16 Dec 2020

%\cite{Dokshitzer:1977sg}
\bibitem{Dokshitzer:1977sg}
Y.~L.~Dokshitzer,
%``Calculation of the Structure Functions for Deep Inelastic Scattering and e+ e- Annihilation by Perturbation Theory in Quantum Chromodynamics.,''
Sov. Phys. JETP \textbf{46}, 641-653 (1977).
%4112 citations counted in INSPIRE as of 16 Dec 2020




%%%%% BFKL
%\cite{Kuraev:1977fs}
\bibitem{Kuraev:1977fs}
E.~A.~Kuraev, L.~N.~Lipatov, V.~S.~Fadin,
%``The Pomeranchuk Singularity in Nonabelian Gauge Theories,''
Sov. Phys. JETP \textbf{45}, 199-204 (1977).
%3325 citations counted in INSPIRE as of 16 Dec 2020

%\cite{Kuraev:1976ge}
\bibitem{Kuraev:1976ge}
E.~A.~Kuraev, L.~N.~Lipatov, V.~S.~Fadin,
%``Multi - Reggeon Processes in the Yang-Mills Theory,''
Sov. Phys. JETP \textbf{44}, 443-450 (1976).
%1668 citations counted in INSPIRE as of 16 Dec 2020

%\cite{Fadin:1975cb}
\bibitem{Fadin:1975cb}
V.~S.~Fadin, E.~A.~Kuraev, L.~N.~Lipatov,
%``On the Pomeranchuk Singularity in Asymptotically Free Theories,''
Phys. Lett. B \textbf{60}, 50-52 (1975).
%!TEX encoding = UTF-8 Unicodedoi:10.1016/0370-2693(75)90524-9
%1237 citations counted in INSPIRE as of 16 Dec 2020

%\cite{Lipatov:1976zz}
\bibitem{Lipatov:1976zz}
L.~N.~Lipatov,
%``Reggeization of the Vector Meson and the Vacuum Singularity in Nonabelian Gauge Theories,''
Sov. J. Nucl. Phys. \textbf{23}, 338-345 (1976).
%1316 citations counted in INSPIRE as of 16 Dec 2020

%\cite{Balitsky:1978ic}
\bibitem{Balitsky:1978ic}
I.~I.~Balitsky and L.~N.~Lipatov,
%``The Pomeranchuk Singularity in Quantum Chromodynamics,''
Sov. J. Nucl. Phys. \textbf{28}, 822-829 (1978)
%3842 citations counted in INSPIRE as of 06 Jul 2022

%\cite{Lipatov:1985uk}
\bibitem{Lipatov:1985uk}
L.~N.~Lipatov,
%``The Bare Pomeron in Quantum Chromodynamics,''
Sov. Phys. JETP \textbf{63}, 904-912 (1986)
LENINGRAD-85-1137.
%933 citations counted in INSPIRE as of 16 Dec 2020


%%%%% CCFM
%\cite{Ciafaloni:1987ur}
\bibitem{Ciafaloni:1987ur}
M.~Ciafaloni,
%``Coherence Effects in Initial Jets at Small q**2 / s,''
Nucl. Phys. B \textbf{296}, 49-74 (1988).
%doi:10.1016/0550-3213(88)90380-X
%802 citations counted in INSPIRE as of 16 Dec 2020

%\cite{Catani:1989yc}
\bibitem{Catani:1989yc}
S.~Catani, F.~Fiorani, G.~Marchesini,
%``QCD Coherence in Initial State Radiation,''
Phys. Lett. B \textbf{234}, 339-345 (1990).
%doi:10.1016/0370-2693(90)91938-8
%649 citations counted in INSPIRE as of 16 Dec 2020



%%%%%LINKED DIPOLE CHAIN MODEL
%\cite{Gustafson:1986db}
\bibitem{Gustafson:1986db}
G.~Gustafson,
%``Dual Description of a Confined Color Field,''
doi:10.1016/0370-2693(86)90622-2.
%325 citations counted in INSPIRE as of 16 Dec 2020

%\cite{Gustafson:1987rq}
\bibitem{Gustafson:1987rq}
G.~Gustafson, U.~Pettersson,
%``Dipole Formulation of QCD Cascades,''
Nucl. Phys. B \textbf{306}, 746-758 (1988). 
%doi:10.1016/0550-3213(88)90441-5
%402 citations counted in INSPIRE as of 16 Dec 2020

%\cite{Andersson:1995ju}
\bibitem{Andersson:1995ju}
B.~Andersson, G.~Gustafson, J.~Samuelsson,
%``The Linked dipole chain model for DIS,''
Nucl. Phys. B \textbf{467}, 443-478 (1996).
%doi:10.1016/0550-3213(96)00114-9
%169 citations counted in INSPIRE as of 16 Dec 2020


%%%%%LUND MODEL
%\cite{Andersson:1998tv}
\bibitem{Andersson:1998tv}
B.~Andersson,
%``The Lund model,''
Camb. Monogr. Part. Phys. Nucl. Phys. Cosmol. \textbf{7}, 1-471 (1997).
%141 citations counted in INSPIRE as of 16 Dec 2020

%\cite{LHCForwardPhysicsWorkingGroup:2016ote}
\bibitem{LHCForwardPhysicsWorkingGroup:2016ote}
K.~Akiba \textit{et al.} [LHC Forward Physics Working Group],
%``LHC Forward Physics,''
J. Phys. G \textbf{43}, 110201 (2016)
doi:10.1088/0954-3899/43/11/110201
[arXiv:1611.05079 [hep-ph]].
%152 citations counted in INSPIRE as of 14 Dec 2021


%%%%% MUELLER-NAVELET
%\cite{Mueller:1986ey}
\bibitem{Mueller:1986ey}
A.~H.~Mueller, H.~Navelet,
%``An Inclusive Minijet Cross-Section and the Bare Pomeron in QCD,''
Nucl. Phys. B \textbf{282}, 727-744 (1987). 
%doi:10.1016/0550-3213(87)90705-X
%468 citations counted in INSPIRE as of 16 Dec 2020


%\cite{deLeon:2020myv}
\bibitem{deLeon:2020myv}
N.~B.~de Le\'on, G.~Chachamis and A.~Sabio Vera,
%``Multiperipheral final states in crowded twin-jet events at the LHC,''
Nucl. Phys. B \textbf{971}, 115518 (2021)
doi:10.1016/j.nuclphysb.2021.115518
[arXiv:2012.09664 [hep-ph]].
%0 citations counted in INSPIRE as of 13 Dec 2021

%%%%% Chew PIGNOTTI
%\cite{Chew:1968fe}
\bibitem{Chew:1968fe}
G.~F.~Chew, A.~Pignotti,
%``MULTIPERIPHERAL BOOTSTRAP MODEL,''
Phys. Rev. \textbf{176}, 2112-2119 (1968).
%doi:10.1103/PhysRev.176.2112
%309 citations counted in INSPIRE as of 16 Dec 2020


%%%%% DeTar
%\cite{Detar:1971qw}
\bibitem{Detar:1971qw}
C.~E.~DeTar,
%``Momentum spectrum of hadronic secondaries in the multiperipheral model,''
Phys. Rev. D \textbf{3}, 128-144 (1971). 
%doi:10.1103/PhysRevD.3.128
%206 citations counted in INSPIRE as of 16 Dec 2020

%\cite{Bzdak:2012tp}
\bibitem{Bzdak:2012tp}
A.~Bzdak, D.~Teaney,
%``Longitudinal fluctuations of the fireball density in heavy-ion collisions,''
Phys. Rev. C \textbf{87} (2013) no.2, 024906.
%doi:10.1103/PhysRevC.87.024906
%[arXiv:1210.1965 [nucl-th]].
%57 citations counted in INSPIRE as of 17 Dec 2020

%%%%%BFKLex
%\cite{Chachamis:2011rw}
\bibitem{Chachamis:2011rw}
G.~Chachamis, M.~Deak, A.~Sabio~Vera, P.~Stephens,
%``A Comparative study of small x Monte Carlos with and without QCD coherence effects,''
Nucl. Phys. B \textbf{849}, 28-44 (2011). 
%doi:10.1016/j.nuclphysb.2011.03.011
%[arXiv:1102.1890 [hep-ph]].
%47 citations counted in INSPIRE as of 17 Dec 2020
%\cite{Chachamis:2011nz}

\bibitem{Chachamis:2011nz}
G.~Chachamis, A.~Sabio Vera,
%``The Colour Octet Representation of the Non-Forward BFKL Green Function,''
Phys. Lett. B \textbf{709}, 301-308 (2012).
%doi:10.1016/j.physletb.2012.02.036
%[arXiv:1112.4162 [hep-th]].
%45 citations counted in INSPIRE as of 17 Dec 2020

%\cite{Chachamis:2012fk}
\bibitem{Chachamis:2012fk}
G.~Chachamis, A.~Sabio~Vera,
%``The NLO N =4 SUSY BFKL Green function in the adjoint representation,''
Phys. Lett. B \textbf{717}, 458-461 (2012). 
%doi:10.1016/j.physletb.2012.09.051
%[arXiv:1206.3140 [hep-th]].
%41 citations counted in INSPIRE as of 17 Dec 2020

%\cite{Chachamis:2012qw}
\bibitem{Chachamis:2012qw}
G.~Chachamis, A.~Sabio Vera, C.~Salas,
%``Bootstrap and momentum transfer dependence in small $x$ evolution equations,''
Phys. Rev. D \textbf{87}, no.1, 016007 (2013).
%doi:10.1103/PhysRevD.87.016007
%[arXiv:1211.6332 [hep-ph]].
%32 citations counted in INSPIRE as of 17 Dec 2020

%\cite{Chachamis:2015zzp}
\bibitem{Chachamis:2015zzp}
G.~Chachamis, A.~Sabio Vera,
%``Monte Carlo study of double logarithms in the small x region,''
Phys. Rev. D \textbf{93}, no.7, 074004 (2016).
%doi:10.1103/PhysRevD.93.074004
%[arXiv:1511.03548 [hep-ph]].
%28 citations counted in INSPIRE as of 17 Dec 2020

%\cite{Chachamis:2015ico}
\bibitem{Chachamis:2015ico}
G.~Chachamis, A.~Sabio Vera,
%``The high-energy radiation pattern from BFKLex with double-log collinear contributions,''
JHEP \textbf{02}, 064 (2016).
%doi:10.1007/JHEP02(2016)064
%[arXiv:1512.03603 [hep-ph]].
%28 citations counted in INSPIRE as of 17 Dec 2020

%\cite{deLeon:2021ecb}
\bibitem{deLeon:2021ecb}
N.~B.~de Le\'on, G.~Chachamis and A.~Sabio Vera,
%``Average minijet rapidity ratios in Mueller\textendash{}Navelet jets,''
Eur. Phys. J. C \textbf{81}, no.11, 1019 (2021)
doi:10.1140/epjc/s10052-021-09811-4
[arXiv:2106.11255 [hep-ph]].
%0 citations counted in INSPIRE as of 13 Dec 2021

%\cite{Cacciari:2011ma}
\bibitem{Cacciari:2011ma}
M.~Cacciari, G.~P.~Salam and G.~Soyez,
%``FastJet User Manual,''
Eur. Phys. J. C \textbf{72}, 1896 (2012)
doi:10.1140/epjc/s10052-012-1896-2
[arXiv:1111.6097 [hep-ph]].
%4124 citations counted in INSPIRE as of 07 Jun 2021

%%%%% EXPERIMENTAL PAPERS
%\cite{Aad:2011jz}
\bibitem{Aad:2011jz}
G.~Aad \textit{et al.} [ATLAS],
%``Measurement of dijet production with a veto on additional central jet activity in $pp$ collisions at $\sqrt{s}=7$ TeV using the ATLAS detector,''
JHEP \textbf{09}, 053 (2011).
%doi:10.1007/JHEP09(2011)053
%[arXiv:1107.1641 [hep-ex]].
%103 citations counted in INSPIRE as of 05 Jan 2021

%\cite{Chatrchyan:2012pb}
\bibitem{Chatrchyan:2012pb}
S.~Chatrchyan \textit{et al.} [CMS],
%``Ratios of dijet production cross sections as a function of the absolute difference in rapidity between jets in proton-proton collisions at $\sqrt{s}=7$ TeV,''
Eur. Phys. J. C \textbf{72}, 2216 (2012).
%doi:10.1140/epjc/s10052-012-2216-6
%[arXiv:1204.0696 [hep-ex]].
%52 citations counted in INSPIRE as of 05 Jan 2021


%\cite{Aad:2014pua}
\bibitem{Aad:2014pua}
G.~Aad \textit{et al.} [ATLAS],
%``Measurements of jet vetoes and azimuthal decorrelations in dijet events produced in $pp$ collisions at $\sqrt{s}=7\,\mathrm{TeV}$ using the ATLAS detector,''
Eur. Phys. J. C \textbf{74}, no.11, 3117 (2014).
%doi:10.1140/epjc/s10052-014-3117-7
%[arXiv:1407.5756 [hep-ex]].
%43 citations counted in INSPIRE as of 05 Jan 2021



%\cite{Khachatryan:2016udy}
\bibitem{Khachatryan:2016udy}
V.~Khachatryan \textit{et al.} [CMS],
%``Azimuthal decorrelation of jets widely separated in rapidity in pp collisions at $ \sqrt{s}=7 $ TeV,''
JHEP \textbf{08}, 139 (2016).
%doi:10.1007/JHEP08(2016)139
%[arXiv:1601.06713 [hep-ex]].
%54 citations counted in INSPIRE as of 28 Dec 2020


\end{thebibliography}
\end{document}